# Electron transport through rectifying self-assembled monolayer diodes on silicon: Fermi level pinning at the molecule-metal interface

*S. Lenfant [1], D. Guerin[1], F. Tran Van[2], C. Chevrot[2], S. Palacin[3],*

*J.P. Bourgoin[4], O. Bouloussa[5], F. Rondelez[5] & D. Vuillaume[1] ***

1) Institut d'Electronique, Micro-électronique et Nanotechnologie – CNRS "Molecular Nanostructures & Devices" group, BP 60069, avenue Poincaré, F-59652 Villeneuve d'Ascq cedex, France.  2) Laboratoire de Physicochimie des Polymères et des Interfaces – Université de Cergy-Pontoise, 5 mail Gay Lussac, F-95031 Cergy-Pontoise, France. 3) Chimie des Surfaces et Interfaces - CEA Saclay, F-91191 Gif sur Yvette cedex, France. 4) Laboratoire d'Electronique Moléculaire – CEA Saclay, F-91191 Gif sur Yvette cedex, France.  5) Laboratoire Physico-chimie Curie - CNRS, Institut Curie, 11 rue Pierre et Marie Curie, F-75231 cedex05, Paris, France.

**\*** Corresponding authors: dominique.vuillaume@iemn.unvi-lille1.fr

**Abstract**

We report the synthesis and characterization of molecular rectifying diodes on silicon using sequential grafting of self-assembled monolayers of alkyl chains bearing a $\pi$ group at their outer end (Si/$\sigma$-$\pi$/metal junctions). We investigate the structure-performance relationships of these molecular devices and we examine to what extent the nature of the $\pi$ end-group (change in the energy position of their molecular orbitals) drives the properties of these molecular diodes. Self-assembled monolayers of alkyl chains (different chain lengths from 6 to 15 methylene groups) functionalized by phenyl, anthracene, pyrene, ethylene dioxy-thiophene, ethylene dioxy-phenyl, thiophene, terthiophene and quaterthiophene are synthesized and characterized by contact angle, ellipsometry, Fourier transform infra-red spectroscopy and atomic force microscopy. We demonstrate that reasonably well-packed monolayers are obtained in all cases. Their electrical properties are assessed by dc current-voltage characteristics and high-frequency (1 MHz) capacitance measurements. For all the $\pi$-groups investigated here, we observe rectification behavior. These results extend our preliminary work using phenyl and thiophene groups (S. Lenfant et al., Nano Letters 3, 741 (2003)).The experimental current-voltage curves are analyzed with a simple analytical model, from which we extract the energy position of the molecular orbital of the $\pi$-group in resonance with the Fermi energy of the electrodes. We report the experimental studies of the band lineup in these silicon/alkyl-$\pi$ conjugated molecule/metal junctions. We conclude that Fermi level pinning at the $\pi$-group/metal interface is mainly responsible for the observed absence of dependence of the rectification effect on the nature of the π-groups, even though they were chosen to have significant variations in their electronic molecular orbitals.

## I. Introduction and motivations

As microelectronic devices approach their technological and physical limits,[1,2] molecular electronics, i.e. the molecule-based information technology at the molecular-scale, becomes more and more investigated and envisioned as a promising candidate for the nanoelectronics of the future. From this respect, supramolecular assembly of organic molecules on solid substrates is a powerful "bottom-up" approach for the fabrication of devices for molecular-scale electronics. A useful method is based on the self-assembly of monolayers of organic molecules on solid substrates (SAM).[3] Many reports in the literature concern devices based on SAMs of thiol-terminated molecules chemisorbed on gold surfaces.[3-5] It is also valuable to develop and investigate molecular-scale devices based on SAMs chemisorbed on semiconductors, especially silicon. Silicon is the most widely used semiconductor in microelectronics and a broad family of organic molecules can be grafted on its surface, which opens the possibility of tailoring the surface (e.g. modifying the surface potential, for instance)[6-8] for new and improved hybrid molecular/silicon devices. Between the end of the silicon road-map and the envisioned advent of fully molecular-scale electronics, there is, for sure, a role to be played by such hybrid-electronic devices.[2,9] Since the first adsorption from solution of alkyltrichlorosilane molecules on a solid substrate (mainly oxidized silicon) introduced by Bigelow, Pickett and Zisman[10] and later developed by Maoz and Sagiv,[11] further detailed studies[12-15] have lead to a better understanding of the basic chemical and thermodynamical mechanisms of this self-assembly process. Since the first study of the electronic properties of alkylsilane monolayers by Mann and Kuhn[16] and Polymeropoulos and Sagiv,[17] SAMs on silicon have been demonstrated as high quality ultra-thin tunnel barriers[18,19] and have been used as gate dielectrics in nanometer-scale transistors.[20-22] SAMs of redox molecules (metallophorphyrines, ferrocene) have also been used as molecular memories[23,24] in hybrid CMOS/molecular DRAM circuits.[25] Molecular resonant tunneling diodes on silicon have been also demonstrated.[26]

Recently,[27] we demonstrated a molecular rectifying junction (MRJ) by attaching a donor group (phenyl or thiophene) to the alkyl spacer chain by a sequential grafting on silicon. We obtained rectification ratios up to 35. We showed that the rectification mechanism is a resonance through the $\pi$ molecular orbital (ascribed to the highest occupied molecular orbital – HOMO) in good agreement with calculations and internal photoemission spectroscopy. This approach allowed us to fabricate molecular rectifying diodes compatible with silicon nanotechnologies for future hybrid circuitries. In this paper, we present a systematic study of the synthesis, the structural and electrical properties of these rectifying molecular diodes, and we extend this work to a large number of chemically different $\pi$ end-groups. The main objectives were to investigate the structure-performance relationships of these molecular devices and to examine the extent to which the nature of the $\pi$ end-group drives their electrical properties. In this respect, SAMs of alkyl chains (chain lengths from 6 to 15 methylene groups) functionalized with a large variety of $\pi$ electron rich chemical groups (phenyl, anthracene, pyrene, ethylene dioxy-thiophene, ethylene dioxy-phenyl, thiophene, terthiophene and quaterthiophene) have been synthesized and characterized by contact angle, ellipsometry, X-ray photoelectron spectroscopy (XPS), Fourier transform infra-red spectroscopy (FTIR) and atomic force microscopy (AFM). Their electrical properties have been tested by dc current-voltage characteristics and high-frequency (1 MHz) capacitance measurements.

We chose a large number of $\pi$-groups having different energy levels of their molecular orbitals in gas-pahse. We started with simple benzyl alcohol and 3-thiophenemethanol. Then, we moved from monomers to oligomers: terthiophene and quaterthiophene and to fused oligoacenes: anthracene and pyrene. Our motivations were to establish a relationship between the electrical properties (electronic structure) of the starting π molecules first in vacuum, then when chemisorbed on the silicon substrate and finally the current-voltage rectification behavior of the Si/molecule/metal junctions. From an engineering point of view, the knowledge of this relationship is mandatory to design MRJ with an electrical behavior suitable for device and circuit applications. For instance, in classical semiconductor p-n junctions the threshold voltage for rectification is adjusted by doping, whereas here it is envisioned to do this using chemistry, by changing the nature of the π-group.

It is known that charge transfer and band lineup (i.e. the energy position of the molecular orbitals (MOs) with respect to the Fermi energy of the electrodes) are the key parameters controlling the electronic properties in molecular devices.[28-31] This question has been theoretically addressed[28-31] in ideal metal-molecule-metal junctions with simple molecules (phenyldithiol, alkyl, alkane and phenyl ethynylene). It is also known that the energy positions of the MOs of the gas-phase molecules and of the molecules in interaction with a surface are quite different. The energy levels of the MOs are broadened due to the hybridization with the delocalized wave functions of the metal electrodes, the energy levels are also shifted under the effects of fractional charge transfer at the interface and the HOMO-LUMO gap can be substantially changed. As a typical example, two-photon photoemission experiments on SAMs of pentafluorothiophenolate chemisorbed on a Cu surface showed that the LUMO is down-shifted by 3.1 eV, the HOMO is up-shifted by 2.6 eV, leading to a HOMO-LUMO gap reduction of 5.6eV, compared to gas-phase molecules.[32] At the metal/organic semiconductor contact (as in organic light emitting diodes), the breakdown of the vacuum alignment rule (Schottky-Mott model) has been a major discovery to explain the electronic properties of these devices,[33,34] and a large number of organic molecules deposited in ultra-high vacuum (UHV) on metal surfaces have been analyzed.[33,34] It was also established that monolayers of molecules bearing a dipole moment can modify the electron affinity of semiconductor surfaces and consequently the metal/semiconductor Schottky barrier height.[35-37] On the contrary, reports on molecular-scale junctions are scarce, probably because of a smaller number of investigated molecules and metal surfaces (much of the works focused on gold surface with linear alkanes and linear π-conjugated oligomers). Moreover, in molecular junctions, the molecules are generally chemisorbed from solution or from gas-phase instead of being UHV deposited.

In this work, we report on a strong metal-induced Fermi level pinning which completely controls the electrical behavior of the molecular rectifying diodes. To our knowledge, this feature has not been yet reported for silicon/SAM/metal junctions. On line with reports quoted above, we report this study to help in the understanding of the current-voltage rectification behavior of these Si/alkyl chain - π group molecule/metal junctions.

## II. Experimental Section

On naturally oxidized silicon wafers, bearing hydroxyl groups, the SAMs are usually prepared from the reaction of molecules bearing a halogen- or alkoxy-silane head-group (Si-$X_3$ with X=Cl or O(CH$_3$)$_3$ or O(C$_2$H$_5$)$_3$...).[3,4] In general, it is desirable to introduce particular functionalities on the surface. Unfortunately, many of them are not compatible with the silane (especially the trichlorosilane) head groups. Therefore, we used a sequential strategy. First, we began with the deposition of SAMs with relatively unreactive end-groups (i.e. relative to the trichlorosilane head group). Second, the end-groups are subsequently modified (sequential grafting). We followed Wasserman et al. who reported on the chemical transformation of vinyl-terminated SAMs into carboxylic acid, alcohol, and bromide end-group SAMs.[38] These groups are prone for further surface modifications by other molecules.

### *II.1. Step 1: Alkyl chain self-assembled monolayer synthesis*

In this work, we used n-alkenyl-trichlorosilane ($SiCl_3–(CH_2)_n–CH=CH_2$) with n=6, 9, 12 and 15. The shorter molecules (n=6 and 9), 7-octadecenyl-trichlorosilane (OETS) and 10-undecenyl trichlorosilane (UETS), respectively, are commercially available (purchased from ABCR and used as received). The longer molecules (n=12 and 15), 13-tetradecenyl-trichlorosilane (TETS) and 16-heptadecenyl-trichlorosilane (HETS), respectively, were synthesized according to the protocol developed by Bonnier[39] using enamine synthesis, or by cross- linking the Grignard reagent of ω-undecenyl bromide with the appropriate dibromoalkane to obtain the bromovinylic chain with the desired length. Installation of the $SiCl_3$ group was achieved by Grignard reaction of the alkenyl bromide with $SiCl_4$ (Figure1). We also prepared SAMs with a methyl-terminated molecule, n-octadecyltrichlorosilane (OTS, $CH_3(CH_2)_{17}SiCl_3$), as a standard reference.

These alkyl chain molecules were chemisorbed on naturally oxidized silicon substrates (1 to 1.5 nm thick $SiO_2$ as measured by ellipsometry, *vide infra*) from a dilute solution ($10^{-2}$-$10^{-3}$ M) in an organic solvent (70/30% v/v of hexane or hexadecane and carbon tetrachloride) using the method introduced by Bigelow, Pickett and Zisman[10] and later developed by Maoz and Sagiv.[11] The silicon substrates (purchased from Siltronix) were degenerate n-type (resistivity of ~ $10^{-3}$ Ω.cm) to avoid any voltage drop in the substrate during electrical measurements. Prior to deposition, the substrate were carefully cleaned by extensive wet cleanings (mainly with a piranha solution: $H_2SO_4$:$H_2O_2$ 2/1 v/v, **caution: piranha solution is exothermic and strongly reacts with organics)** and dry cleanings by combining ultraviolet irradiation and ozone atmosphere. The cleaned substrates were dipped into the freshly prepared solution and the chemical reaction was allowed to proceed to completion. Typical reaction times were 90-120 minutes. We deposited HETS at 20 °C, TETS at 8 °C, UETS at 2 °C and OETS at -20 °C while the corresponding critical temperatures ($T_C$) for optimum deposition (i.e. to form a densely-packed and well-ordered monolayer) are ~26, ~16, ~6, and ~-10 °C, respectively. These critical temperatures were extrapolated from Brzoska et al.[12,13] It was assumed that the critical temperatures for the vinyl-terminated alkyl chains were the same as for the methyl-terminated molecules used in Brzoska's work. To control the solution temperature during the silanization reaction, the glass beaker was placed onto a cold plate (temperature controlled at ± 1 °C). To avoid any condensation of water when working at low temperature near the dewpoint of the atmosphere surrounding the reaction bath, we worked in a glove box, purged and maintained under a dry nitrogen flow (relative humidity around 15%). The next step consisted in modifying the vinyl end-groups by oxidation in $KMnO_4/NaIO_4/K_2CO_3$/deionized $H_2O$ (10/10/10/70 % v/v) to obtain the COOH terminated monolayers.[38] The oxidation time was 24h at room temperature. For all of these chemical steps, we worked in a dry nitrogen-purged glove box installed in a class 10000 clean room (temperature and relative humidity well controlled at 20 °C and 40%, respectively).[40]

### II.2. Step 2: Synthesis and grafting of the π end-groups

Finally, we grafted conjugated moieties onto the previously formed SAMs using esterification reactions between the -COOH end-groups and several π-conjugated molecules bearing an alcohol group (R-OH). We started with simple benzyl alcohol and 3-thiophenemethanol (Figures 2.a and 2.b, respectively).[41] Then, we moved from monomers to oligomers: terthiophene and quaterthiophene (Figures 2.g and 2.h) and to fused oligoacenes: anthracene and pyrene (Figures 2.c and 2d).[42] Finally, we also used phenyl and thiophene derivatives substituted with ethylene-dioxy (EDBM and EDTM, Figures 2.e and 2.f).

***Bis (5,5''-(2-hydroxy 2-methylethyl))-2,2':5',2''-terthiophene***. (3T for short, (**g**) in Fig. 2). 2,2':5',2''-terthiophene was first synthesized in diethyl ether using the well known Grignard reaction from 2,5 dibromothiophene and 2 bromothiophene in the presence of Nickel (II) catalyst (Nidpppcl$_2$).[43] mp(°C) 89. [1]H NMR (CDCl$_3$, ppm) 7.2 ( H5,5''); 7.18 ( H3,3''); 7.06 ( H3',4'); 7.0 ( H4,4'' ). 5g (20.1 mmoles) of 2,2' :5',2''-terthiophene were solubilized in dry diethylether and 25.2 mL of butyllithium (2.5M in hexane) were added. Stirring was maintained for further 30 mn at room temperature and then 10 mL (143 mmoles) of propyleneoxide were added drop wise. The resulting mixture was maintained under stirring for 3 h, and then hydrolysed with 30 mL of water. After filtration and extraction of the organic phase, the crude product was recristallized in toluene. 2.2 g of brown crystals were obtained (yield 30%). mp(°C) 96.5. [1]H NMR (CDCl$_3$, ppm) 6.98 (H3',H4', H3, H3''); 6.75 (d, H4,4''); 4.02 (m, 2H, CH$_2$**CH**OHCH$_3$); 2.92 (m, 2H, **CH$_2$**CHOHCH$_3$ ); 1.72 (d, 2H, CH$_2$CHO**H**CH$_3$); 1.28 (d, 6H, CH$_2$CHOHC**H$_3$**).

***2,3-dihydrothieno[3,4-b]-1,4-dioxin-2-yl-methanol***  (or ethylene-dioxy-thiophene methanol - EDTM for short, (**f**) in figure 2) was synthesized in 6 steps using a previously described procedure.[44-46]

***2-hydroxymethyl-1,4-benzodioxan***. (or ethylene-dioxy-benzyl methanol - EDBM for short, (**e**) in figure 2). 2.2g (20 mmoles) of catechol was dissolved in 250 mL of boiling ethanol. Epibromhydrin (2.5 ml, 30 mmoles) and potassium carbonate (0.55 g, 4 mmoles) dissolved in 150 mL of water were then added. The mixture was heated at reflux for 1 h. Additional amounts of K$_2$CO$_3$ (0.3 g) and epibromhydrin (1 ml) were added. After refluxing for 72 h, the solution was cooled and poured into 100 mL of acidified water (5% HCl) . After extraction with chloroform and drying over magnesium sulfate the solvent was removed by evaporation under vacuum. Recristallization in ether gives 2.1 g of white crystals. (yield 63%). mp(°C) 88. [1]H NMR (CDCl$_3$, ppm) 6.87 (m, 4H); 4.26 (m, 2H); 4.1(m, 1H); 3.84(m, 2H); 1.91 (t, 1H).

The esterification was carried out for 120 h at room temperature (always in the class 10000 clean-room and in the glove-box or in a dry nitrogen purged and sealed vessel) in the presence of DCCI (dicyclohexylcarbodiimide) to enhance the reaction yield (benzyl alcohol was used pure, the other compounds were dissolved in xylene at $10^{-2}$ M). The solution was renewed after 60 h to minimize its contamination and aging. An esterification at shorter time led to worse quality SAMs as inferred from structural characterization (contact angle, ellipsometry, FTIR and AFM). Figure 3 shows a schematic representation of these σ-π SAMs for each of the π-groups.

### II.3. Wettability measurements

The contact angles of sessile drops of test liquids were measured using a remote computer-controlled goniometer system (DIGIDROP by GBX, France). A drop of deionized water (18 MΩ.cm) or hexadecane (in the range 1-10 μL) was deposited on the surface and the projected image was acquired and stored by the remote computer. Contact angles were then extracted by contrast contour image analysis software. These static angles were determined 5 s after application of the drop. These measurements were carried out in a clean room (class 1000), in which the relative humidity (40%) and the temperature (20 °C) are well controlled. The accuracy of those measurements was ± 2°.

### II.4. Thickness and dielectric constant measurements

The SAM thicknesses were measured by ellipsometry with a PLASMOS SD2300 instrument at 6328 Å. We used a value of n=1.50 for the SAM refractive index at 6328

Å to calculate the thickness. Usual values in the literature are in the range 1.45-1.50.[3] We also used a spectroscopic ellipsometer UVISEL (by Jobin Yvon) equipped with a DeltaPsi 2 data analysis software. The system acquired a spectrum ranging from 2 to 4.5 eV (corresponding to 300 to 750 nm) with 0.05 eV (or 7.5 nm) intervals. Data were taken using an angle of incidence of 70° and the compensator was set at 45°. Data were fitted by regression analysis to a film-on-substrate model described by their thickness and their complex refractive indices. The optical parameters of the naturally oxidized substrate were independently determined by measuring a bare wafer rigorously cleaned by the same surface cleaning process (*vide supra*). This oxide layer thickness was found to be in the range 10-15 Å for all the wafers used in this work. We compared the measured data with the simulated data to determine this thickness. The simulated data were obtained with a 2 layers model: silicon substrate/silicon oxide. We used for Si and silicon oxide, the optical properties (complex refractive index for each wavelength) from the software library. After the monolayer deposition, we used a 3 layers model: silicon substrate/silicon oxide/organic monolayer. To determine the monolayer thickness we fixed the oxide thickness at the previously measured value, for silicon and oxide we again used the optical properties from the software library, and for the monolayer we fixed the refractive index at 1.50. The accuracy of the SAM ellipsometry thickness measurement is estimated to be ± 2 Å.

We also combined high-frequency (1 MHz) capacitance (*vide infra*) and ellipsometry to determine both the dielectric constant and the thickness of the SAM's. The SAM capacitance is given by $C_{SAM} = \varepsilon_{SAM}\varepsilon_0 / d_{SAM}$ and the ellipsometry optical thickness $K = n.d_{SAM}$ where $\varepsilon_0$ is the vacuum permittivity, $\varepsilon_{SAM}$ the relative SAM permittivity, $d_{SAM}$ the SAM thickness and n the SAM refractive index. Assuming that $\varepsilon_{SAM} = n^2$, we calculated $\varepsilon_{SAM}$, n and $d_{SAM}$ using $d_{SAM} = \left( \varepsilon_0 K^2 \middle/ C_{SAM} \right)^{1/3}$ and $n = \left( KC_{SAM} \middle/ \varepsilon_0 \right)^{1/3}$. This neglects possible dipolar contribution to the dielectric function. This assumption has been validated a posteriori by the fact that the measured thicknesses are in close agreement with theoretical ones (see section III.2).

### II.5. Fourier transform infrared spectroscopy

Fourier transform infrared spectroscopy (FTIR) measurements of the monolayers were done with a Perkin Elmer Spectrum 2000 system, equipped with a liquid nitrogen cooled MCT detector. We used internal reflection infrared spectroscopy (known as attenuated total reflection -ATR). We used a silicon ATR crystal (10 mm x 5 mm x 1.5 mm, faces cut at 45°). All measurements were made after purging the sample chamber for 30 min with dry $N_2$. Spectra were recorded at 4 cm$^{-1}$ resolution, using a strong apodization, and 200 scans were averaged to increase the signal-to-noise ratio. Background spectra were recorded on a freshly cleaned ATR crystal before each monolayer deposition.

### II.6. Atomic force microscopy

We used Atomic Force Microscopy (AFM) to image the surface morphology of the SAMs. We used a Nanoscope III (Digital Instruments) system in the tapping mode in air and at room temperature with a silicon tip. All images (512 x 512 pixels) were taken at the scanning rate of 2 - 2.44 Hz. Surface regions from 50 x 50 nm to 5 x 5 μm were imaged.

### II.7. Electrical measurements

For the capacitance and conductivity measurements, we formed the silicon/SAM/metal (SSM) heterostructures by evaporating metal (aluminum) through a shadow mask (electrode area: $10^{-2}$ mm$^2$). To avoid contaminating the surface during the metallization, an ultra-high vacuum (UHV) e-beam evaporation system was used. It was checked that a $10^{-8}$ Torr vacuum is innocuous for the SAMs. We have shown previously that such SAMs are thermally stable up to ~ 350 °C in vacuum,[47] nevertheless during the evaporation, the sample temperature was maintained below 50 °C. To avoid damage of the monolayer during the deposition of the first monolayers of the evaporated metal atoms, we used a low evaporation rate (1-5 Å/s) and a large distance between the sample and the crucible of metal (~ 70 cm). The electrode thicknesses were in the range 200-500 nm. More than 20 SSM devices were measured for each combination of alkyl chain lengths and $\pi$ end-groups. The success rate for forming working junction was about 50-70% (ratio of non short-circuited devices over total measured ones). Aluminum (instead of Au) was chosen to avoid any rectification effect coming from the difference in the work functions of the two electrodes ($W_M \approx 4.2$ eV for Al and $W_{Si} \approx \chi_{Si} = 4.1$ eV for n$^+$-type Si, $\chi_{Si}$ is the electron affinity) since it is well known that a larger current is obtained when a positive bias is applied to the electrode with the smallest work function.[48] This effect was observed through metal/SAM/metal junction with Au and Ti electrodes.[49]

The SSM structures were mounted onto a wafer chuck with silver paste to insure good electrical contact with the silicon back-side. The electrode was contacted by precision micromanipulators. Electrical transport through the SAMs was determined by measuring the current density versus the applied dc voltage with an Agilent 4140B pico-ammeter. We used a low speed step-like voltage ramp (step voltage 10 mV, duration 1.5 s) to avoid transient effects due to displacement current since the SSM junctions mainly act as capacitors. Capacitances were measured at 1 MHz (ac signal amplitude 20 mV$_{eff}$) by an Agilent 4274A LCR-meter. In both cases, voltages were applied to the metal counter-electrode, the silicon substrate being grounded. The measurements were done at room temperature and in the ambient atmosphere. The SAM capacitance C$_{SAM}$ is deduced from the measured capacitance C$_{meas}$ taking into account the capacitance of oxides (C$_{ox}$) and the capacitance for the silicon substrate (C$_{sc}$): $1/C_{meas} = 1/C_{sc} + 1/C_{ox} + 1/C_{SAM}$. We modeled the degenerate semiconductor by its Debye capacitance, $C_{sc} = C_{Debye} = \varepsilon_0 \varepsilon_{sc} / \lambda_{Debye}$ with ε$_{sc}$ the dielectric constant of the semiconductor (11.9 for silicon) and λ$_{Debye}$ the extrinsic Debye length. This latter is given by $\left( \varepsilon_0 \varepsilon_{sc} kT / q^2 N_D \right)^{1/2}$ with k the Boltzmann constant, q the elementary electron charge, T the temperature and N$_D$ the doping concentration in the semiconductor (here ≈ $10^{19}$ cm$^{-3}$). We get C$_{debye}$ ≈ 8 μF/cm$^2$. The oxide capacitance $C_{ox} = \varepsilon_0 \varepsilon_{ox} / d_{ox}$ was calculated knowing the measured native oxide thickness (ellipsometry) and taking ε$_{ox}$=3.9.

### III. Results and discussion

#### III.1. Contact angles

Figure 4 summarizes the measured evolution of the water contact angles for the different chemical functionalities of the SAMs. Table 1 gives the water and hexadecane contact angles measured on the alkyl chain SAMs after and before the oxidation reaction. For the OTS monolayer used as standard reference, the contact angles are 108°±2° for water and 43°±2° for hexadecane as expected for a densely-packed methyl terminated SAM.[3,12,13] The vinyl-terminated SAMs are clearly hydrophobic with 97°<θ$_{H2O}$<105°. The oxidation reaction makes them more hydrophilic with $20° < \theta_{H2O} < 56°$, depending on the length of the alkyl chains. After the esterification with any of the $\pi$ groups, the water contact angles are always in the range 70 – 80°. Our contact angles for the vinyl-terminated monolayers value are in good agreement with earlier reports on similar systems.[3,50] For instance, Steel et al.[50] have reported $\theta_{H2O}=108°\pm3°$ and $\theta_{HD}=39°\pm3°$ for a vinyl-terminated alkylthiol SAM on gold (chain length of 9 methylene groups). Both the water and hexadecane contact angle values compare well with those obtained in this work for the longest chain length (HETS). Our hexadecane contact angles decrease with decreasing chain lengths from 15 to 6 methylene groups, becoming too low and not measurable (<10°) for the two shortest chain lengths. This reflects a more disordered SAM for the short chain length molecules (see section 3.3 on FTIR results) and the fact that the underneath hydrophilic $SiO_2$ surface can contribute to the wetting properties for such a short chain.[51] After oxidation, the water contact angle should have been equal to 0° for a nearly ideal, 100% covered, COOH-terminated monolayer.[52] We measured higher values indicating a partial COOH surface coverage. Assuming that the monolayer surface is made of a mixture of vinyl and acid carboxylic moieties, and using the Cassie law[53], we estimate the oxidation reaction yield.

$$\cos \theta_{vin/carb} = r_{vin}\cos \theta_{vin} + r_{carb}\cos \theta_{carb} \qquad (1)$$

where $\theta_{vin/carb}$ is the measured contact angle after oxidation, $r_{vin}$ is the relative surface coverage by vinyl groups and, $r_{carb}$ with the relative surface coverage by COOH groups ($r_{vin} + r_{carb} =1$), $\theta_{vin}$ and $\theta_{carb}$ are the water contact angles for an ideal, fully covered surface, by vinyl and COOH groups, respectively. We used 105° and 0°, respectively. We obtain the oxidation yields $\eta_{ox} = r_{carb}$ (if we assume that $r_{vin}=1$ for the monolayer before the oxidation) of 68±5%, 63±5%, 74±5%, 95±5% for the OETS, UETS, TETS and HETS monolayers, respectively (Table 1). These yields are in agreement with data reported by Wasserman et al.,[38] these authors have estimated the yield of this chemical surface modification in the range 70% to 90%. However, we notice that the shorter the chain length, the lower the water contact angle and the deduced oxidation yield. This may be explained by an increase of the disorder when decreasing the chain length (see section 3.3 for FTIR confirmation), and thus we overestimate the oxidation yield because $r_{vin}$ is < 1 for the SAM before oxidation. After esterification, the contact angles are in agreement with other reports on similar systems (monolayer functionalized by thiophene[54], phenyl[50]). These values are also in agreement with the one expected if the aromatic groups are densely-packed and if they preferentially expose their edges to the probe liquid.[55]

### III.2. Ellipsometry

Table 2 summarizes the thicknesses of the SAMs and gives a comparison with the expected value for a "near-perfect" densely-packed SAM. The expected thickness corresponds to the length of the molecule, as given by PM3 geometry optimization with the CS-MOPAC software[56], and assuming that the main axis of the molecule is perpendicular to the surface substrate. We observe a relatively good agreement between the measured and the expected values. The differences are always lower than 5 Å and may be ascribed to irreproducibility and sample-to-sample variations of the native oxide (1 to 1.5 nm). This indicates a reasonably good packing of these SAMs, especially after the grafting of the π end-groups.

In the case of the vinyl-terminated monolayers, we measured 12±2 Å, 17±2 Å, 18±2 Å and 26±2 Å for the OETS, UETS, TETS and HETS monolayer, respectively. These values are in agreement with the general expression of the molecule length obtained by Wasserman et al.[38] for a methyl-terminated monolayer containing *n* methylene units ($SiCl_3$-$(CH_2)_n$-$CH_3$) : $d_{SAM}$ = 1.26 n + 4.78  (in Å). This relation has been determined for methyl-terminated monolayers, while our monolayers are vinyl-terminated. But the length difference between these two end-groups is very low (inferior to 0.2 Å).[57] The OTS monolayer thickness is equal to 26±2 Å in perfect agreement with the above relation and previous results.[3,38]

### II.3 FTIR

For alkylsilane monolayers, the frequency and width of the C-H stretching bands are indicative of the degree of order of the alkyl chains within the monolayers.[3] The $CH_2$ vibration peaks for the reference OTS and the precursor HETS monolayers are $\nu_a$ = 2918±1 $cm^{-1}$ (antisymmetric) and $\nu_s$ = 2850±1 $cm^{-1}$ (symmetric) (Figure 5 and Table 3). These values are the fingerprint of a dense and well-ordered monolayer,[3,12,14,15] The positions of these peaks are strictly similar to those of $CH_3$ terminated chains (OTS) used as reference. We inferred from these values that the alkyl chains in these SAMs are in their all-trans conformation, nearly perpendicular to the substrate (tilt angle < 10°) as expected for a densely-packed molecular architecture in the SAM. Amplitudes (areas) of these peaks scale linearly with chain length.

For the monolayers with shorter alkyl chains, the peaks shift to higher wave numbers (Fig. 5 and Table 3). For example, in the case of the OETS monolayer, the peak positions are at 2927 and 2858 $cm^{-1}$ for the antisymmetric and the symmetric mode, respectively. This behavior has been observed in the case of alkylthiol grafted on gold[3] or platinum[58] substrates. This peak shifts are due to a decrease in the Van der Waals interactions between neighbor molecules (they increase with the length of the alkyl chain), leading to an increase in the disorder into the monolayer when the chains are shorter.

The C-H peak positions (symmetric and antisymmetric) after oxidation (i.e. for the COOH terminated SAM) and after esterification (with the π-group) are the same as for the starting vinyl-terminated SAMs (Table 3). The peak amplitude and width are also not changed by these successive chemical processes. We conclude that the chemical functionalization of the starting vinyl-terminated SAMs by oxidation and esterification with the aromatic moieties does not degrade the molecular organization of the alkyl chains in the monolayers.

In the region of π bonds, we observed the C=O stretching vibration at 1717 cm$^{-1}$ (Figure 6) corresponding to the C=O bond in a COOH group.[59] After the esterification with aromatic moieties, the stretching C=O modes shifts to 1735 cm$^{-1}$. This higher value is in agreement with literature values for the C=O vibration mode in an ester bond.[59] In addition, the C=C stretching band at ~1651 cm$^{-1}$ is clearly observed after esterification, which evidences that the aromatic moieties are effectively attached to the previously formed SAMs.

### III.4. AFM

Figure 7 shows typical AFM images for the bare substrate, the HETS SAM and the same HETS SAM functionalized by the pyrene group. Both SAMs are homogeneous; we did not observe holes (at the resolution of the AFM, ~few tens of nanometers). However, the rms roughness increases, especially after the grafting of the pyrene group. We found a rms roughness ~0.11 nm for the bare substrate, ~0.16 nm for the HETS SAM and ~0.22 nm for the HETS pyrene. This increase may be due to the larger size of the pyrene compared to the alkyl chain and/or to a more disordered organization of the pyrene groups than for the alkyl ones. Note that both SAMs (HETS and HETS-pyrene) display some hollow (darker area) with a depth of about 0.5-0.7 nm. These "defects" may explain the dispersion observed in the amplitude of current-voltage curves and the fact that a fraction of the alkyl SAM were short-circuited (~30%) and that a fraction of the alkyl-π group SAM showed a weak rectification ratio (< 2) or no rectification at all (see below for details and ref. [27]).

## IV. Electrical properties

### IV.1. Capacitance measurements

To determine the impact of the π group on the dielectric permittivity, we measured the capacitance of the monolayers. Capacitance may change for two reasons: i) an increase in thickness and ii) a change of the dielectric permittivity when anchoring π moieties with higher dipole moment than the pure alkyl chain (almost non polar). Combining capacitance and ellipsometry measurements (see section 2.4) allows us to determine both the dielectric constant and the thickness of the end-group functionalized SAM's. From the capacitance measurements, we can extract the ratio $\varepsilon_{SAM}/d_{SAM}$ where $\varepsilon_{SAM}$ is the SAM dielectric constant and $d_{SAM}$ the SAM thickness, and from the ellipsometry we can extract the product $n.d_{SAM}$, where $n$ is the optical refraction index and $d_{SAM}$ the thickness. Assuming that $\varepsilon_{SAM}=n^2$, we combined both results to calculate $\varepsilon_{SAM}$, $n$ and $d_{SAM}$ (Table 4). The refractive index ($n$) values for all the monolayers are around 1.5. Similar value of $n$ for both alkyl and alkyl-π SAMs is expected because the π-groups are not highly polar. This result confirms the choice of 1.5 for the refractive index made in the section 2.4 for the ellipsometry measurements, and the similar values of $\varepsilon$ for the alkyl chain and the π-group in eq. (3) – see below. The thicknesses deduced by combining ellipsometry and capacitance measurements are in agreement with the expected ones and with the thicknesses measured by ellipsometry alone.

### IV.1. Conductivity measurements

Figure 8 shows typical current-density vs. voltage curves (J-V) for several π-terminated SAMs made with HETS (Figs 8-a and 8-b) and OETS (Fig. 8-c) as the alkyl spacer. The J-V curves for alkylsilane SAM (without π group) are taken as the reference (figure 8-d). The J-V curves for the $CH_3$-terminated (OTS) and vinyl-terminated (OETS, TETS and HETS) SAMs are symmetric (see figure 8-d, the case of OTS). We have also checked that the introduction of the polar ester group is not responsible for the rectification behavior in the π-terminated SAMs. The figure 8-d shows the J-V for the COOH-terminated SAM (as obtained after step 1 described above and based on a pristine HETS SAM). In the COOH-terminated case, the shape of the J-V is less linear and a small asymmetry is observed with a slightly higher current at positive bias (a ratio ~1.35 at $\left| 1V \right|$ ), i.e. in the opposite way compare to the π-terminated SAMs. In all the other cases, with the π moieties at the end, we observed a rectification behavior i.e. a higher current density at -1V than at 1V. To further analyze this rectification effect, we compared the dispersion for the current densities at -1 V and 1 V (see the example for HETS/phenyl in figure 9). The observed distributions are fitted by two Gaussian curves that are not at the same position; the current density at -1V is shifted to higher values (by almost a factor 10 in that case). This statistical analysis clearly reveals the rectification behavior in spite of the inherent current dispersion already observed with these SAMs. The electrical properties for all the monolayers studied in this work are presented in table 5. For every π-functionalized monolayer, we define a rectification ratio (RR for short) equal to the average current density at -1 V (in absolute value) divided by the average current density at 1V (RR = $|J_{-1V}| / J_{1V}$). The average values of RR are in the range 3 to 13 (table 5) for the functionalized monolayers, with the highest RR value of 37 for the OETS/thiophene monolayer. We did not observe a significant variation of RR with the chemical nature of the end-groups nor any correlation with the alkyl spacer length (the sample to sample variations of the RR values are too large with relative scattering of about 50% for a given end-group and alkyl chain length).

Figure 10 is a schematic representation of the energy diagram of the silicon (covered by native $SiO_2$) /alkyl-π group/metal junction: (a) at 0 V; (b) at a negative bias on the metal electrode for which a resonant electron transfer can occur through the π-level from the metal to the empty states of the Si CB. We have $W_M \approx 4.2$ eV for Al and $W_{Si} \approx \chi_{Si} = 4.1$ eV for n$^+$-type Si, $\chi_{Si}$ is the electron affinity. σ, π and σ*, π* are the HOMO and LUMO of the alkyl chain and π-group, respectively. σ* is about 4.2 eV above the Si CB and σ is about 4.2 eV below the Si-VB.[19,27] The positions of the CB and VB of the ultra-thin (~1 nm) native $SiO_2$ are not exactly given, but they are likely below the LUMO and HOMO of the alkyl chain, respectively. For instance, values as low as ~ 1 eV have been reported for the CB of such an ultra-thin native oxide (see compiled data in Fig. 3 of reference [60]). The positions of the π and π* levels and $E_0$ should depends on the nature of the π-groups. We also notice that the role played by the native oxide as a tunnel barrier is negligible (at first order) compared to that of the alkyl chain monolayer. Comparing the current densities in Al/Si, Al/native $SiO_2$/Si, Al/C18/Si and Al/C18/native $SiO_2$/Si (where C18 stands for a monolayer of a 18 carbon atoms alkyl chain), we have measured[18,60,61] (at 0.5 V) ~ $10^2$ A/cm$^2$, ~ 50 A/cm$^2$ and ~ $10^{-7} - 10^{-8}$ A/cm$^2$ for the two latter junctions, respectively. This justifies that we neglect the native oxide in analyzing the J-V curves of the present devices. The rectification behavior of these σ-π SAMs is due to the resonant tunneling through one of the MO's of the π group (Fig. 10) due to the geometrical asymmetry (the π group is closer to the metal electrode than the silicon) and to the energy asymmetry in the positions of the MO's with respect to the Fermi level of the electrodes.[62,63] The rectification effect arises for a negative bias applied on the Al electrode because the energy difference between the silicon Fermi energy (pinned at the conduction band - CB - for the degenerate Si) and the HOMO (π orbital) is lower than that with the LUMO (π* orbital). If we assume that the π-end group is almost at the Al electrode potential (since this group is in close contact with the electrode) and that a large part of the potential drop takes place in the alkyl chains, the threshold (in absolute value) required to have a resonance is lower for a negative bias than for a positive bias. As a consequence, the J-V curves can be fitted by a one-level model in which the conduction is dominated by the charge transport through a single energy level located at $E_0$ below ($E_0<0$) the electrode Fermi energy. This allows an experimental determination of its energy position. The current density is given by,[64,65]

$$J = \frac{2J_0}{\pi}\left\{\tan^{-1}\left[\theta\left(|E_0|+\eta eV\right)\right] - \tan^{-1}\left[\theta\left(|E_0|-(1-\eta)eV\right)\right]\right\} \quad (2)$$

where V is the applied potential on the metal electrode (Si is grounded), e is the electron charge, η is the fraction of the potential across the π moiety, $J_0$ is the saturation current and θ is an electrode/molecule coupling parameter. The η value was estimated using a simple dielectric model where the σ and π parts of the SAM have a thicknesses $d_\sigma$ and $d_\pi$ and a dielectric constants $\varepsilon_\sigma$ and $\varepsilon_\pi$, respectively.[62,66]

$$\eta = 1 - \frac{1}{2}\frac{1}{1+\dfrac{\varepsilon_\pi d_\sigma}{\varepsilon_\sigma d_\pi}} \quad (3)$$

In practice, the dielectric constant of the $\pi$-groups used in this work and the alkyl chains are almost similar (see next section),[67] so eq. (3) reduces to $\eta=(d_\sigma+0.5d_\pi)/(d_\sigma+d_\pi)$, which simply represents the relative position of the center of gravity of the $\pi$ moiety measured from the Si substrate. Taking the thicknesses derived from the ellipsometry measurements of the SAM before and after esterification of the $\pi$ end-group (see section 3.2), the typical values of ~0.83, ~0.87 and ~0.9 are obtained for the SAMs based on OETS, TETS and HETS, respectively. As shown in Fig. 8, the measured J-V curves can be satisfactorily adjusted with this formula ($R^2 \geq 0.97$). The solid lines in figure 8 are the best fits obtained with these equations and the values of the adjustable parameters $J_0$, $E_0$ and $\theta$ are given in table 5. We have then looked after a relationship between the intrinsic electronic structure of the $\pi$ group (their MOs in vacuum), the electronic structure of the Si/$\sigma$-$\pi$/metal junction (MOs of the $\pi$ group embedded in the junction) and the MRJ experimental behavior. For this ,we have compared the HOMO energy of the $\pi$ group in vacuum (PM3 calculation), the HOMO energy of the $\pi$ group as deduced from gas-phase ionization potential, and the experimental HOMO position in the junctions respective to the silicon Fermi energy (i.e. $E_0$ deduced from the fit of the analytical current density equation as shown above). These values are compared in table 5 and figure 11. It is clear that the experimentally determined HOMO level of the $\pi$ groups in the MRJ markedly differs from the single molecule values (both PM3 and IP). As explained in the introduction, there are many reasons explaining these differences: molecule/surface interactions leading to charge transfer and interface dipole formation, polarization energy in the solid-state that move the HOMO (LUMO) upwards (downwards, respectively) with respect to the gas-phase values, intrinsic dipole of the molecule itself. For instance, PM3 calculations show that all the molecules used in this work have a dipole moment of about 2-4 D along their long axis with the positive charge at the Si side and the negative one at the ester side. As a consequence, this dipole shifts the MO's of the $\pi$-group upwards by a quantity $\Phi = \mu_D / \varepsilon A_{mol}$ (assuming the molecules with their long axis normal to the surface) where $\mu_D$ is the moment of the molecule, $\varepsilon$ the dielectric constant and $A_{mol}$ the area per molecule. Assuming a relative dielectric constant of 3 and an average $A_{mol}$ of 30 - 50 $\text{Å}^2$ (depending on the molecule), we get $\Phi$ in the range 1 to 2 eV. These values are consistent with several experimental determination for SAM on both metal and Si surfaces,[68,69] but are not sufficient to explain the observed differences. More detailed calculations (both ab-initio and semi-empirical) to calculate the full electronic structure of the Si/$\sigma$-$\pi$/metal junctions are in progress and will be reported elsewhere.[70] A striking experimental feature is that the HOMO energy positions in the MRJ are almost the same whatever the $\pi$ molecules. They vary from 0.71 to 0.82 eV below the Si conduction (Fig. 11-a), or in other words, the metal Fermi energy level is pinned at about 0.71 - 0.82 eV above the $\pi$ HOMO. This is a very small variation compared to the one expected, $\sim 1.5$ eV, from the gas-phase ionization and PM3 levels (Fig. 11-a). However, plotting the HOMO levels arbitrarily normalized to the level for the phenyl end-group (Fig. 11-b), we show that the relative trends for the variations of the HOMO levels versus the nature of the $\pi$-groups are conserved in the molecular junction, the amplitude of these variations being screened in the junction. This behavior is the fingerprint of Fermi level pinning at the metal/molecule interface. A usual way to quantify the Fermi level pinning consist in calculating the so-called interface slope parameter, $S = \left| \dfrac{dE_F}{dW_M} \right|$, where $W_M$ is the metal work function and $E_F$ the position of the Fermi level with respect to one of the molecular orbitals. Here, since we used only one metal and various organic molecules, one can equivalently determine, $S = \left| \dfrac{dE_F}{dI_P} \right|$, where $I_P$ is the ionization potential of the

molecule. S=1 corresponds to the Schottky-Mott model[71,72] and S=0 to the Bardeen[73]

model. In the former case, the energetics of the interface is strictly dictated by the

difference in the work function of the two materials, the latter case assumes that a high

density of interface states pins the position of the Fermi level whatever the nature of the

metal electrode. From the plot of $E_F$ vs $I_P$ (Fig. 12) we deduced an average slope

S=0.025±0.02. This slope is related to the density of metal/organic interface states at the

Fermi level, $D_{it}(E_F)$ by[74] $D_{it} = \varepsilon_0 \varepsilon_i (1-S)/e\delta S$, with $\varepsilon_0$ the vacuum dielectric constant, $\varepsilon_i$ dielectric constant of the organic/metal interface region, $\delta$ its thickness and e the electron charge. We do not know the actual interface dipole in our devices[75] but we can nevertheless try to get an estimate of $D_{it}$. Let us consider $\delta$ being about 5 Å and assuming $\varepsilon_i \sim 2.5 - 3$ as in the SAM, we get $D_{it}(E_F)$ in the range $\sim 10^{15}$ cm$^{-2}$eV$^{-1}$.

A possible origin of the observed metal Fermi pinning is the existence of metal-

induced gap states (MIGS)[76] or the creation of chemically-induced gap states (CIGS) at

the metal/organic interface due to the possible reaction of aluminum with the $\pi$-

conjugated moieties.[33] Recently, MIGS at metal/organic interfaces has been theoretically and experimentally studied in PTCDA/Au.[76] The creation of MIGS results in a pinning of the metal Fermi level very near the charge neutrality level (CNL), i.e.

the energy position for which the total charge integrated over the band-gap density is null. In our case, another likely origin of the metal Fermi pinning is the creation a CIGS at the metal/organic interface due to the possible chemical reaction of aluminum with the π-conjugated moieties.[33] Typical examples of this behavior are the Alq$_3$/Mg and Alq$_3$/Al interfaces.[77,78] The chemical reactivity of vapor-deposited Al on SAMs of alkyl chains bearing various end-groups has been widely studied.[79-83] It was shown that Al reacts with oxygen-based terminal groups such as –COOH, -CO$_2$CH$_3$, -OH, -OCH$_3$ forming organoaluminum complexes. However, reports on vapor deposition of metals on SAMs bearing a conjugated end groups are scarce. Ahn and Whitten[84] have observed a strong chemical interaction between vapor-deposited Al and a thiophene-terminated SAM which appears as a metal-induced, low-binding energy components on the X-ray photoemission spectroscopy S$_{2p}$ and C$_{1s}$ main peaks. Similarly, de Boer and coworkers reported (infra red spectroscopy) that Al atoms reacted with the conjugated backbone of thiol-oligophenyl SAMs on gold. Thus, it is likely that Al chemically reacts with any of the 8 π-groups used in this work, or even with the ester group if some Al atoms penetrate into the SAMs. Detailed theoretical calculations are in progress to determine the electronic structure of the whole Si/molecule/metal junction which requires an exact treatment of the metal/organic interface dipole.[76,85] Finally, figure 13 gives the MRJ threshold voltage V$_T$, which is defined as the intercept between a linear fit of the current at high negative voltages and the x-axis. All these values are summarized in table 5. As a consequence of the pinning of the Fermi level, V$_T$ is independent of the nature of the π group. A negative bias of $V_T \approx -E_0 / e\eta$ (i.e. ~ - E$_0$/e since η is ~1, see above) is necessary to line up the HOMO on resonance with the CB of the n$^+$-doped silicon. The actual values V$_T$~ - 0.65 to - 0.7 V are quite consistent with this explanation. A further improvement would be to chemically tune the rectification behavior of the molecular diode. It requires decoupling the $\pi$ group from the metal electrode. This could be achieved by introducing a short alkyl spacer chain between the $\pi$ group and the top electrode. For instance, it has been calculated that rectification will persist if the ratio of the number of carbon atoms in the lower and upper alkyl spacer chains is larger than 2.[86]

## V. Conclusion

We have demonstrated that SAMs containing $\pi$-groups (phenyl, anthracene, pyrene, ethylene dioxy-thiophene, ethylene dioxy-phenyl, thiophene, terthiophene and quaterthiophene) can be obtained by sequential grafting of alkyl chains (different chain lengths from 6 to 15 methylene groups) which are functionalized in a second step. Such SAMs are reasonably well structured at a macro and microscopic scale, as can be seen by contact angle, ellipsometry, IR spectroscopy and AFM measurementrs. For all the $\pi$-groups investigated here, we have observed a rectification behavior in their current-density vs. voltage characterisitcs, which extends our preliminary work using phenyl and thiophene groups.[27] A simple analytical model was fitted on the experimental current-voltage curves to determine the position of the $\pi$-group molecular orbitals with respect to the electronic structures of the silicon substrate and the metal top electrode. The electronic structure of these molecular rectifying junctions can be calculated using a self-consistent tight-binding method. Comparing with the experimental data allows us to conclude that Fermi level pinning at the $\pi$-group/metal interface is mainly responsible for the observed behavior. It also explains why the rectification effect does not depend on the nature of the π-groups, albeit they have been chosen to have significant variations in their electronic molecular orbitals in vacuum.

**Acknowledgements**. We thank A. Leroy and A. Fatorrini (IEMN clean room facilities) for making the numerous metallizations. We acknowledge the gift of the 4T-methanol by A. Yassar (ITODYS). We acknowledge financial support from the CNRS and the ministry of research through the "AC nanosciences" program (projects "DIODEMOL" and "MONOFET"), and from the "Institut de Recherche sur les Composants logiciels et matériels pour l'Informatique et la Communication Avancée" (IRCICA).

**Table 1.** Contact angles with water ($\theta_{H2O}$) and hexadecane ($\theta_{HD}$) for each monolayer before and after oxidation. The yield of this reaction ($\eta_{ox}$) for each molecule is calculated by the Cassie Law (see text for details).

| | vinyl-terminated monolayer | | COOH-terminated monolayer | | |
|------|------|------|------|------|------|
| | $\theta_{H2O}$ | $\theta_{HD}$ | $\theta_{H2O}$ | $\theta_{HD}$ | $\eta_{ox}$ |
| HETS | 105±2° | 37±2° | 20±2° | <10° | 95±5% |
| TETS | 102±2° | 18±2° | 47±2° | <10° | 74±5% |
| UETS | 100±2° | <10° | 56±2° | <10° | 63±5% |
| OETS | 97±2° | <10° | 50±2° | <10° | 68±5% |

**Table 2.** Thickness measured by ellipsometry for the different alkyl chain lengths (HETS, TETS, UETS and OETS) and the different π-end groups. The measured values (meas.) are compared with the calculated length (calc.) of the molecule (PM3 calculations). Error bar is ± 2 Å for all measurements.

| | HETS | | TETS | | UETS | | OETS | |
|---|---|---|---|---|---|---|---|---|
| | meas. (Å) | calc. (Å) | meas. (Å) | calc. (Å) | meas. (Å) | calc. (Å) | meas. (Å) | calc. (Å) |
| Vinyl | 25.8 | 25.0 | 18.0 | 21.2 | 17.4 | 15.9 | 12.0 | 12.2 |
| COOH | 24.6 | 23.8 | 18.2 | 20.0 | 16.7 | 15.7 | 11.0 | 12.0 |
| Phenyl | 31.8 | 28.9 | 21.6 | 25.2 | 20.5 | 20.8 | 19.0 | 17.2 |
| Thiophene | 30.0 | 28.2 | 23.0 | 24.5 | 24.7 | 20.1 | 14.5 | 16.9 |
| Anthracene | 30.2 | 29.3 | | | | | | |
| Pyrene | 31.2 | 32.3 | 26.5 | 28.6 | | | | |
| EDBM | 29.5 | 31.8 | | | | | | |
| EDTM | 27.8 | 31.3 | | | | | | |
| 3T | | | | | 30.1 | 31.7 | 28.0 | 28.0 |
| 4T | | | | | 29.7 | 36.2 | 33.0 | 32.5 |

**Table 3.** Peak positions of the symmetric ($\upsilon_s$) and the antisymmetric ($\upsilon_{as}$) C-H stretching bands for the different monolayers.

|  | Vinyl-terminated monolayer | | COOH-terminated monolayer | | $\pi$–terminated monolayer | |
|---|---|---|---|---|---|---|
|  | $\nu_{as}$ (cm$^{-1}$) | $\nu_s$ (cm$^{-1}$) | $\nu_{as}$ (cm$^{-1}$) | $\nu_s$ (cm$^{-1}$) | $\nu_{as}$ (cm$^{-1}$) | $\nu_s$ (cm$^{-1}$) |
| OTS | 2918±1 | 2850±1 |  |  |  |  |
| HETS | 2918±1 | 2850±1 | 2918±1 | 2850±1 | 2918±1 | 2850±1 |
| TETS | 2924±1 | 2853±1 | 2924±1 | 2853±1 | 2924±1 | 2853±1 |
| UETS | 2926±1 | 2856±1 | 2926±1 | 2856±1 | 2926±1 | 2856±1 |
| OETS | 2927±1 | 2858±1 | 2927±1 | 2858±1 | 2927±1 | 2858±1 |

**Table 4.**     Summary of the capacitance measured at 1MHz for the different monolayers. Taking into account the oxide capacitance, the Debye capacitance of the semiconductor, we deduce the capacitance of the monolayer. Coupling these measurements with ellipsometry (from Table 2) give us the dielectric constant and the thickness of the monolayers (see text or details)

| monolayer | $C_{SAM}$ ($\mu F/cm^2$) | $\varepsilon_{SAM}$ | n | $d_{SAM}$ (Å) | $d_{th}$ (Å) |
|---|---|---|---|---|---|
| TETS/phenyl | 1.00±0.20 | 2.25±0.02 | 1.5±0.1 | 21±3 | 25 |
| TETS/thiophene | 1.00±0.50 | 2.35±0.03 | 1.5±0.3 | 22±3 | 25 |
| TETS/pyrene | 1.10±0.70 | 2.75±0.09 | 1.7±0.5 | 23±3 | 28 |
| HETS | 0.80±0.10 | 2.33±0.04 | 1.5±0.2 | 25±2 | 25 |
| HETS/phenyl | 0.66±0.04 | 2.30±0.03 | 1.5±0.2 | 31±2 | 29 |
| HETS/thiophene | 0.66±0.02 | 2.21±0.02 | 1.5±0.1 | 30±2 | 28 |
| HETS/anthracene | 0.94±0.01 | 2.80±0.01 | 1.7±0.1 | 27±3 | 29 |
| HETS/pyrene | 0.59±0.09 | 2.10±0.05 | 1.5±0.2 | 32±2 | 32 |
| HETS/EDBM | 0.61±0.03 | 2.06±0.02 | 1.4±0.2 | 31±3 | 32 |
| HETS/EDTM | 0.77±0.01 | 2.32±0.01 | 1.5±0.1 | 27±3 | 31 |
| OTS | 1.06±0.05 | 2.7±0.03 | 1.6±0.1 | 22±2 | 26 |

**Table 5.**    Average rectification ratio (RR=│J(-1V)│/J(+1V)), saturation current density ($J_0$), experimental position of HOMO level ($E_0$) respective to the Si-CB, coupling parameter ($\Theta$), theoretical HOMO in vacuum (PM3 calculations), gas-phase ionization potential (eV) and molecular diode threshold voltage for the various π-end groups. In all cases, the energy reference is taken at the Si-CB, we have subtracted the Si work function, $W_{Si}$ = 4.1 eV from the HOMO-PM3 and gas-phase usually referenced to the vacuum level.  (a) alkyl chain = HETS, (b) = OETS.

| π-group | RR | $J_0$ (A.cm$^{-2}$) | $E_0$ (eV) | θ (eV$^{-1}$) | HOMO PM3 (eV) | IP (eV) | $V_T$ (V) |
|---|---|---|---|---|---|---|---|
| Benzene | 4.8±3.3 (a) | $1.3\times10^{-5}$ | -0.82±0.16 | 5.0±2.9 | -5.65 | -5.1[87] | -0.64±0.07 |
| Thiophene | 13.3±13 (b) | $3.3\times10^{-4}$ | -0.80±0.14 | 7.5±2.8 | -5.44 | -4.7[87] | -0.65±0.07 |
| | 2.7±1.4 (a) | $1.8\times10^{-5}$ | | | | | |
| EDBM | 4.7±2.9 (a) | $7.8\times10^{-5}$ | -0.75±0.03 | 8.5±0.5 | -4.92 | -3.6[88] | -0.68±0.01 |
| EDTM | 6.7±2 (a) | $6.6\times10^{-6}$ | -0.73±0.03 | 7.4±2.7 | -4.88 | -4.2[88] | -0.68±0.04 |
| pyrene | 9.4±3.8 (a) | $9.1\times10^{-6}$ | -0.71±0.04 | 10.1±1.5 | -4.15 | -3.45[87] | -0.68±0.02 |
| anthracene | 8.1±2.9 (a) | $9.4\times10^{-6}$ | -0.75±0.09 | 10.4±1.2 | -4.15 | -3.25[87] | -0.7±0.02 |
| 3T | 3.6±1.4 (b) | $4.7\times10^{-3}$ | -0.81±0.1 | 8.6±1.6 | -4.58 | -3.3[89] | -0.68±0.05 |
| 4T | 4.6±0.2 (b) | $9.4\times10^{-3}$ | -0.77±0.07 | 9.2±0.3 | -4.56 | -3.18[89] | -0.7±0.02 |

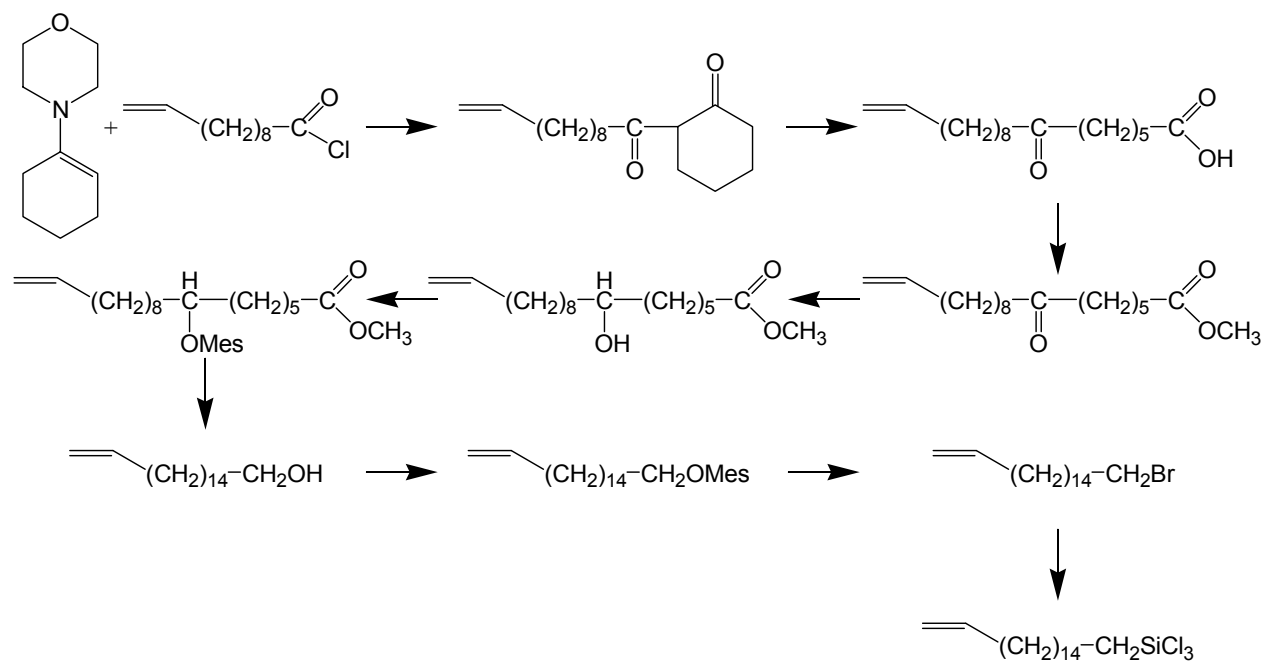

**Figure 1.** Outline of the synthesis for the longer molecule: 16-heptadecenyl-trichlorosilane (or HETS) according to Bonnier.[39]

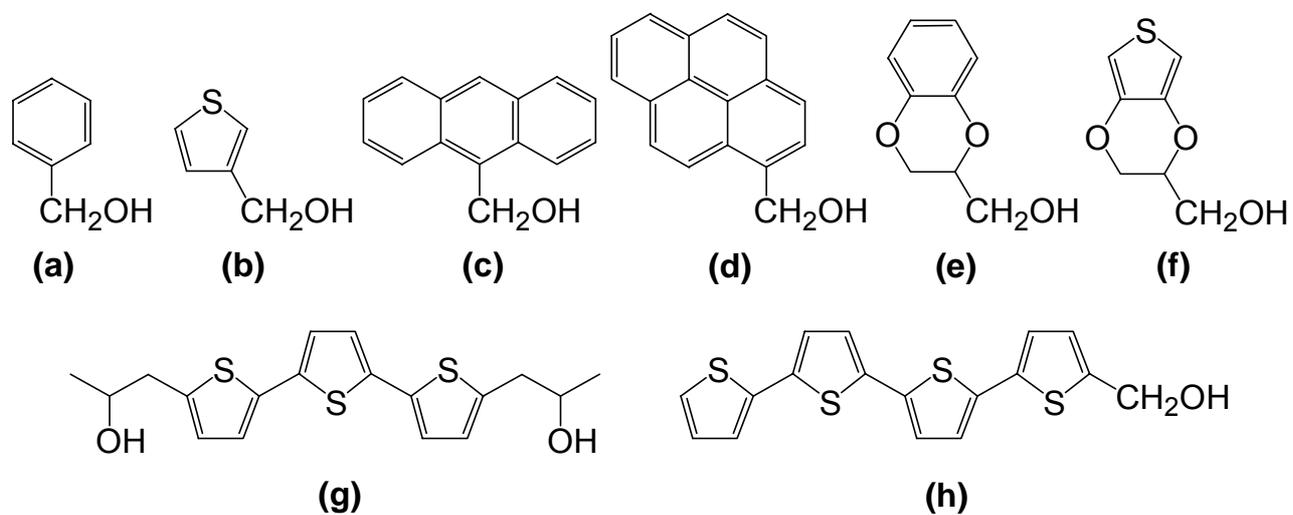

**Figure 2.** Chemical structures of compounds used to functionalize the SAMs benzyl alcohol (a), 3-thiophenemethanol (b), 9-anthracenemethanol (c), 1-pyrenemethanol (d), 3,4-ethylenedioxybenzene methanol (EDBM) (e), 3,4 - ethylenedioxythiophene methanol (EDTM) (f), bis (2-hydroxypropyl) terthiophene (g), and hydroxymethyl quaterthiophene (h).

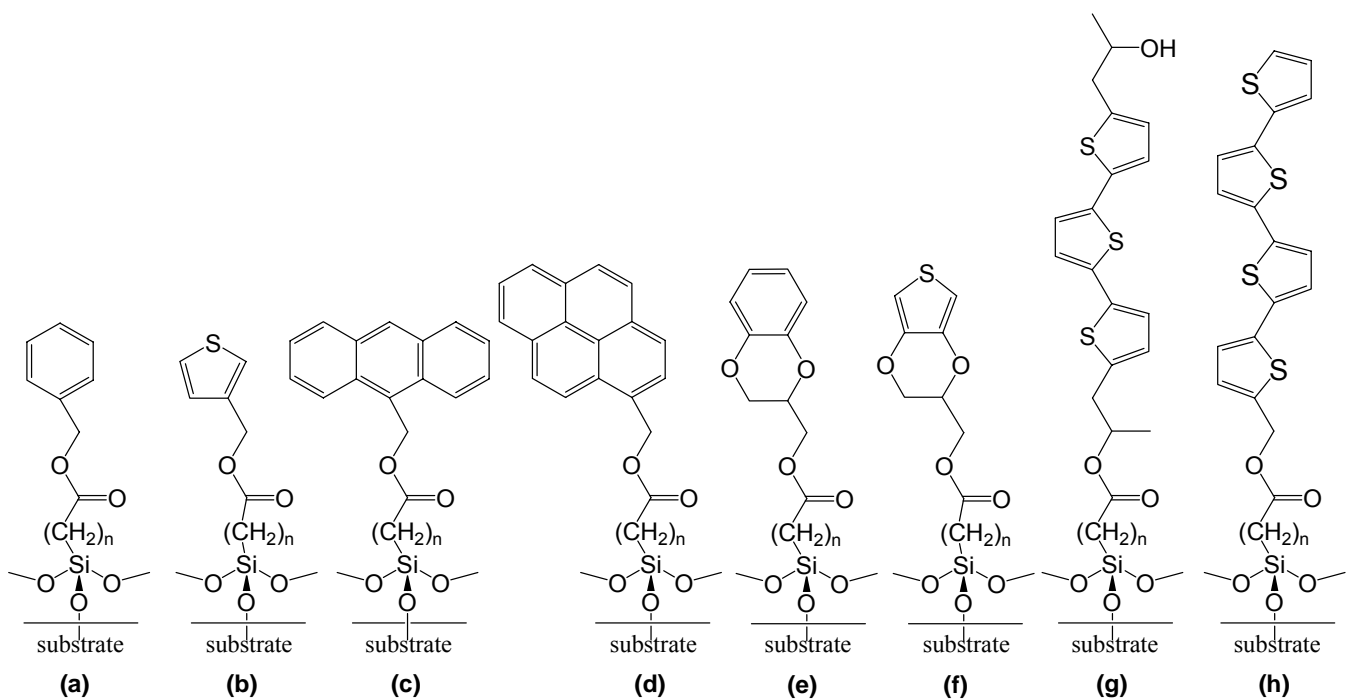

**Figure 3.** Schematic view of the final SAMs with height different aromatic molecules studied in this work: benzyl alcohol (a), 3-thiophenemethanol (b), 9-anthracenemethanol (c), 1-pyrenemethanol (d), 3,4-ethylenedioxybenzene methanol (EDBM) (e), 3,4 -ethylenedioxythiophene methanol (EDTM) (f), bis (2-hydroxypropyl) terthiophene (g), and hydroxymethyl quaterthiophene (h).

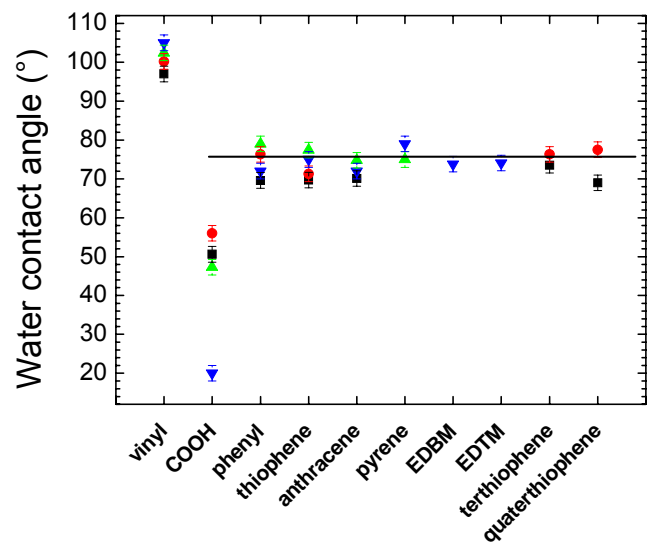

**Figure 4.** Evolution of the water contact angles for the different chemical functionalities of the SAMs, for each molecule used for the silanization: HETS (▼), TETS (▲), UETS (●), OETS (■).

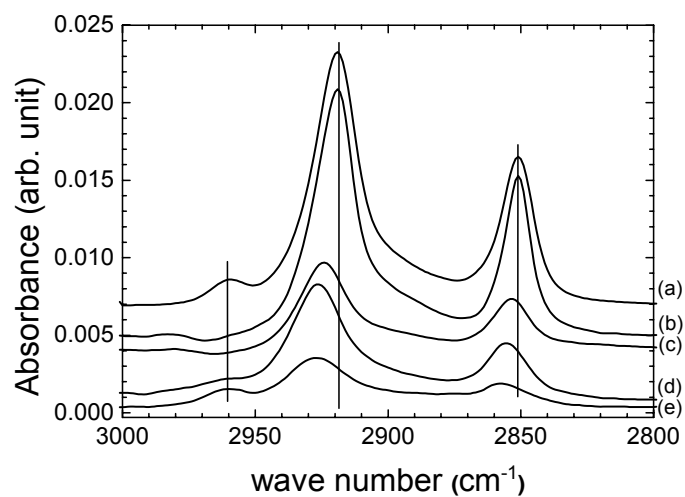

**Figure 5.** ATR-FTIR spectra in the C-H region obtained on a monolayers of OTS (a), HETS (b), TETS (c), UETS (d) et OETS (e). The intensities are given in transmission absorbance units. The vertical lines at 2960, 2918 and 2850 cm$^{-1}$ are provided as a guide to the eye.

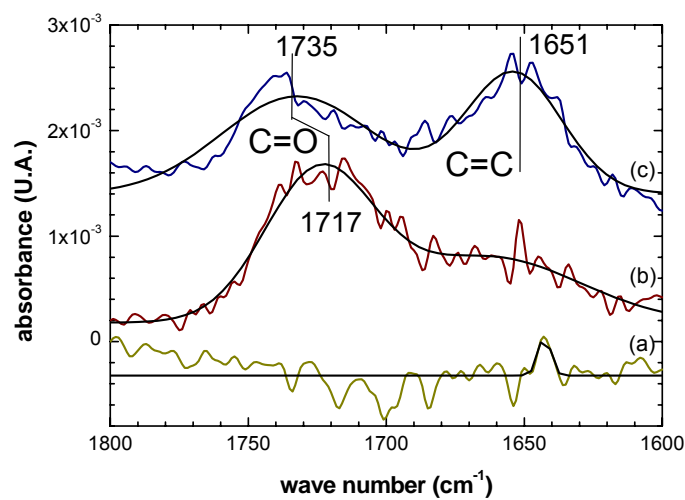

**Figure 6.** ATR-FTIR spectra in the C=C and C=O region obtained on a vinyl- (a), COOH- (b), and anthracene- (c) terminated monolayers. Here we present a HETS monolayer functionalized by anthracene as an example. The vertical lines are provided as a guide to the eye.

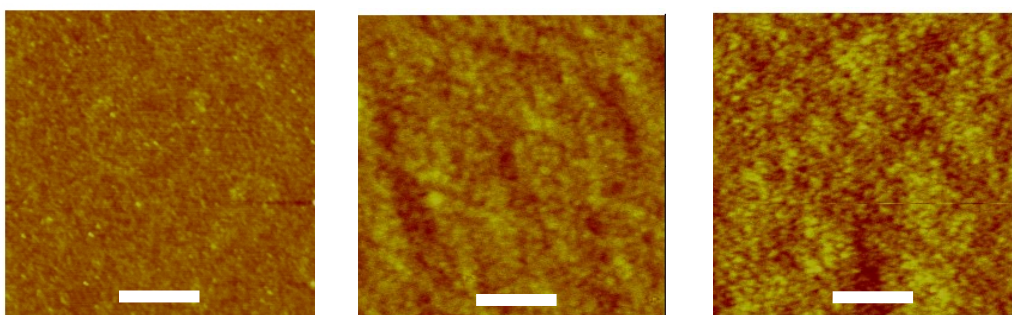

**Figure 7.**　Typical AFM images (1μm x 1μm) for the bare substrate (left), HETS SAM (middle) and HETS-pyrene SAM (right). Scale bar is 250 nm.

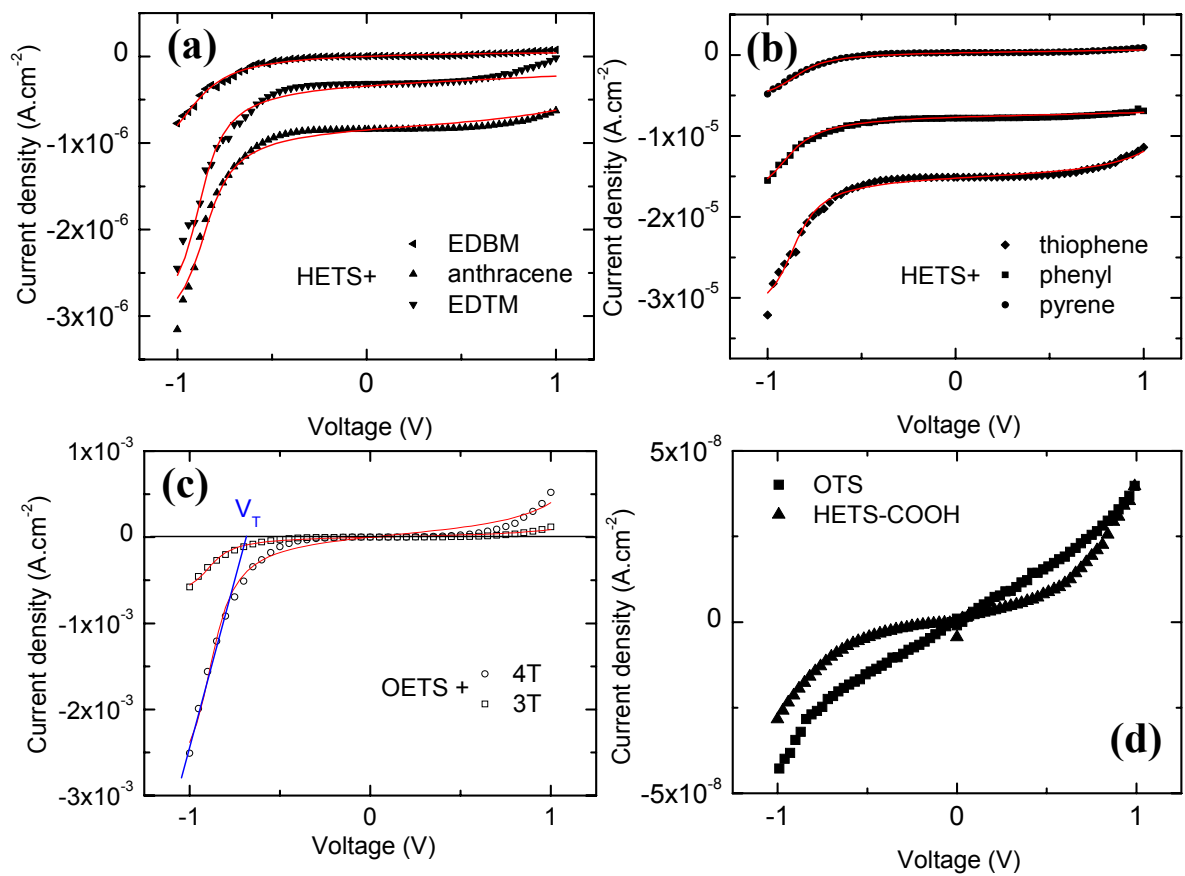

**Figure 8.** Typical current density - voltage characteristics of the Si/σ-π/Al junctions : **(a)** HETS/EDBM (◄), HETS/anthracene (▲), HETS/EDTM (▼) ; **(b)** HETS/phenyl (■), HETS/pyrene (●), HETS/thiophene (♦); **(c)** OETS/3T (□), OETS/4T (○). Some curves are vertically shifted for clarity. Red solid lines are fits by equation (2). For the OETS/4T, we show the determination of threshold voltage for rectification, $V_T$, which is defined as the intercept between a linear fit (dotted line) of the current at high negative voltages and the x-axis. **(d)** The current-voltage characteristic for the OTS and COOH-terminated HETS (i.e. without π end group) are presented as reference.

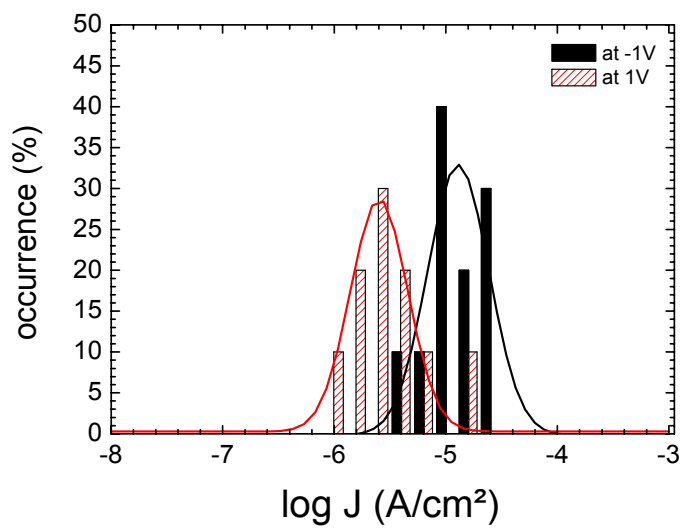

**Figure 9.** Example of the dispersion for the current density for the HETS/phenyl SAM (10 measured samples). We compare the dispersion for the current densities at -1V and 1V. For each monolayer, the center of the Gaussian and the FWHM (Full Width Half Maximum) give respectively the average value and the statistical dispersion for the current density.

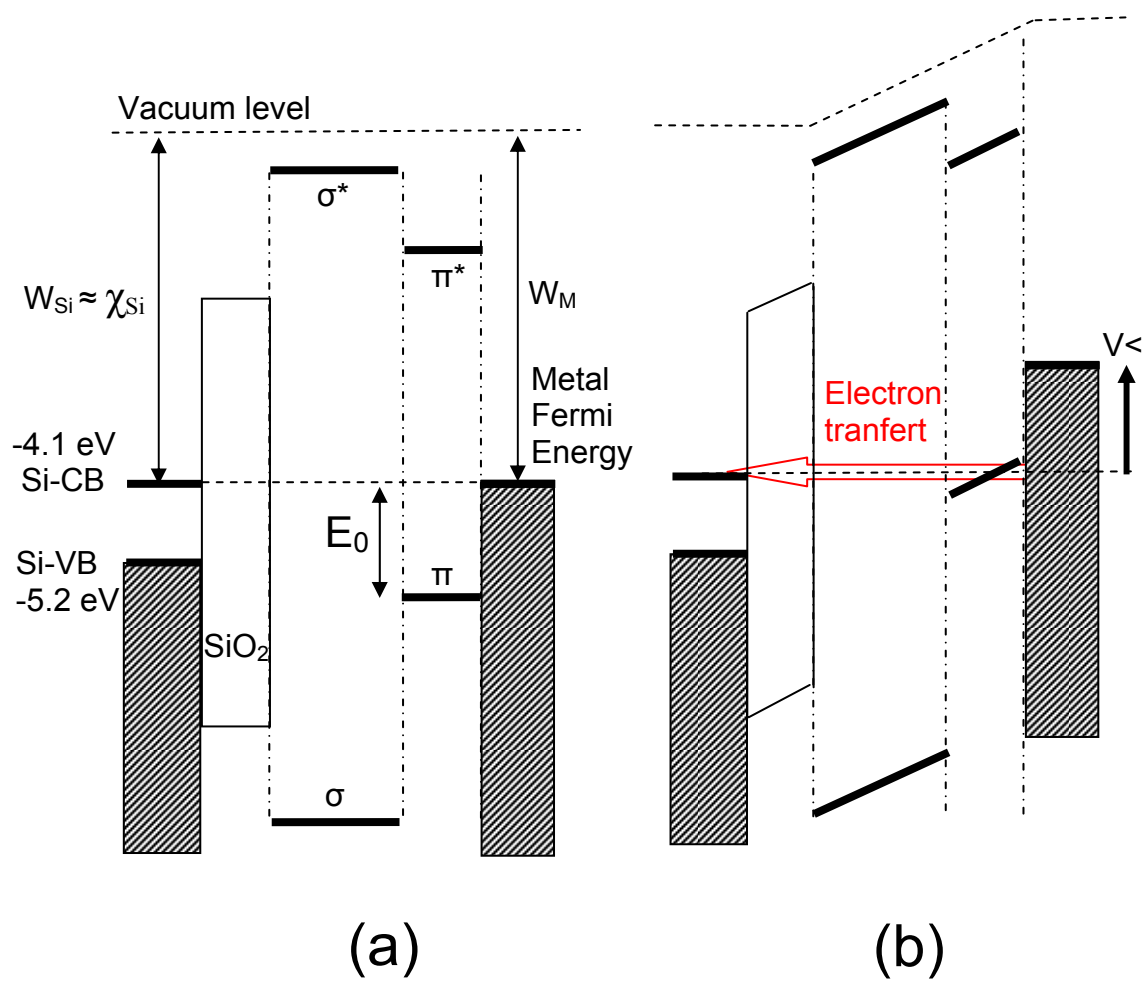

**Figure 10.** Schematic representation of the energy diagram of the silicon (covered by native SiO$_2$) /alkyl-$\pi$ group/metal junction: (a) At 0 V; (b) at a negative bias on the metal electrode for which a resonant electron transfer can occur through the $\pi$-level from the metal to the empty states of the Si CB, for simplicity we assumed a linear potential drop through the SAM.

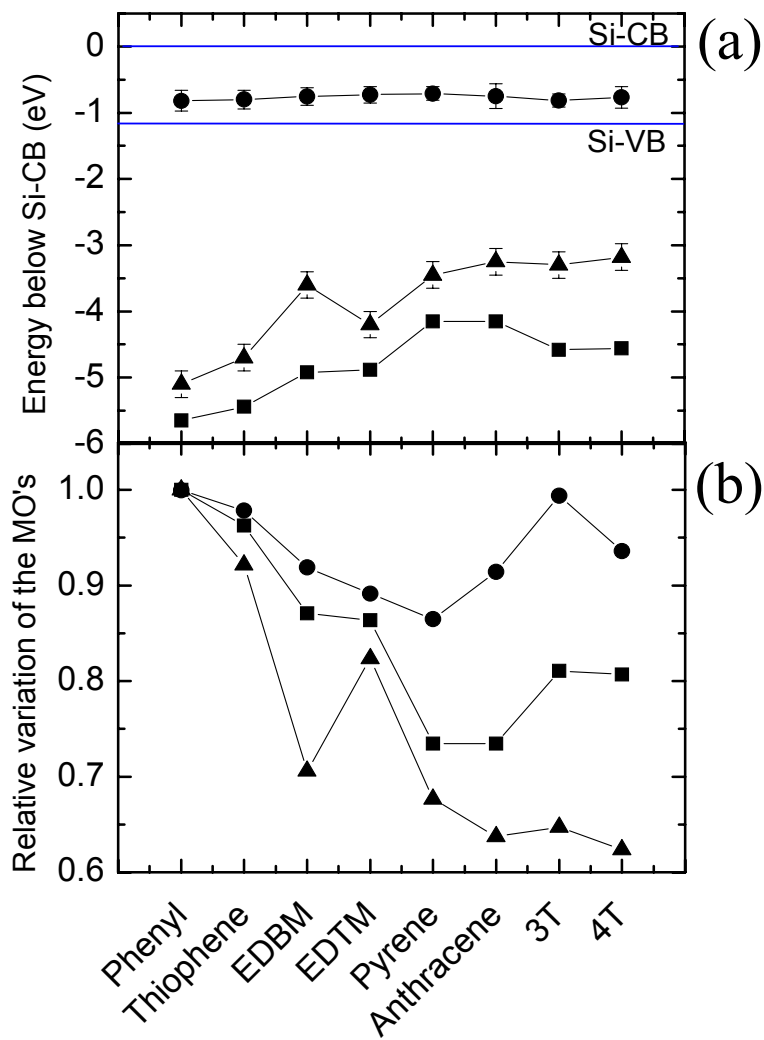

**Figure 11.** **(a)** Energy position (with respect to the Si-CB) for the different π groups : (●) for the chemisorbed SAM on Si as determined from the present experiments; (■) PM3 calculations for a single molecule in vacuum; (▲) experimental gas-phase ionization potential (see references in table 5). **(b)** Same data normalized to the corresponding value for the phenyl group.

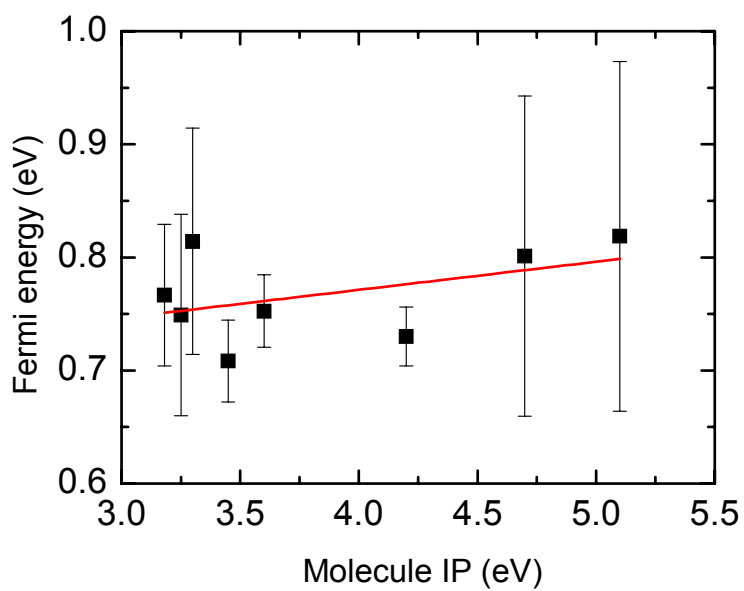

**Figure 12.** Metal Fermi energy position (with respect to HOMO of the π-group) versus the gas-phase IP of the π groups (data from table 5 and figure 11-a). The plotted value correspond to $-E_0$ where $E_0$ is determined from the fit of eq (1) on the J-V curves. The gas-phase IP is the absolute value with respect to Si-CB level. The line is a linear fit with a slope of ~0.025 ± 0.02.

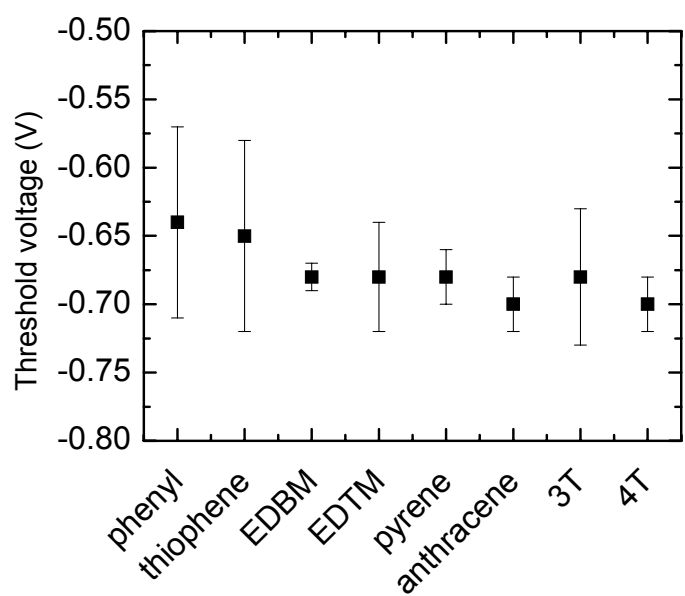

**Figure 13.** Threshold voltage for rectification, $V_T$, as a function of the nature of the $\pi$ end group. $V_T$ is defined as the intercept between a linear fit (dotted line in figure 8) of the current at high negative voltages and the x-axis (see figure 8). The error bars are the FWHM (Full Width Half Maximum) of the data statistical distribution obtained on a large number (>20) of devices for each $\pi$ group.